\documentclass[runningheads,a4paper]{llncs}

\usepackage{amssymb}
\usepackage{amsmath}
\usepackage{multicol}
\usepackage{fixltx2e}

\usepackage{graphicx}
\usepackage{caption}
\usepackage{subcaption}
\usepackage{cite}
\usepackage{url}
\urldef{\mailsa}\path|{andreea.buga, t.nemes}@cdcc.faw.jku.at|

\usepackage{listings}

\lstdefinelanguage{AsmetaL} 
{morekeywords={par, endpar, if, endif, then, else, seq, endseq, signature, definitions, asm, module, import, function, domain, main, rule, macro, invariant, over, choose, forall, exists, let, endlet, with, ifnone, abstract, default, init, do, agent, dynamic, controlled, monitored, in, out, static, extend, derived, subsetof, switch, case, endswitch, enum, CTLSPEC, LTLSPEC, JUSTICE},
	sensitive=true, morecomment=[l]{//}, morecomment=[s]{/*}{*/},
	morecomment=[l][\color{white}\tiny]{'},
	morestring=[b]",tabsize=2, columns=fullflexible, basicstyle=\sffamily, captionpos=b}
\lstdefinelanguage{Avalla} 
{morekeywords={scenario, load, invariant, set, exec, step, until, check},
	sensitive=true, morecomment=[l]{//}, morecomment=[s]{/*}{*/},
	morecomment=[l][\color{white}\tiny]{'},
	morestring=[b]",tabsize=2,columns=fullflexible}
\usepackage[hidelinks]{hyperref}
\usepackage{combelow}
\usepackage{newunicodechar}
\newunicodechar{ș}{\cb{s}}

\usepackage{booktabs}

\usepackage[ paperheight  =297mm,paperwidth   =210mm,  
             layoutheight =200mm,layoutwidth  =120mm,
             layoutvoffset= 41.9mm,layouthoffset= 45mm,
             centering,
             margin=0pt, includeheadfoot,
             footskip=5mm,
           ]{geometry}

    {\large\bfseries Acknowledgement%
    \par\medskip\normalfont\normalsize}%
    {}%
\begin{document}
\mainmatter  

\title{Addressing Client Needs for Cloud Computing using Formal Foundations\thanks{The research reported in this chapter has been supported by the Christian-Doppler Society in the frame of the Christian-Doppler Laboratory for Client-Centric Cloud Computing and by the Austrian Ministry for Transport, Innovation and Technology, the Federal Ministry of Science, Research and Economy, and the Province of Upper Austria in the frame of the COMET center SCCH.}}

\titlerunning{Addressing Client Needs for Cloud Computing using Formal Foundations}

%
%
\author{Andreea Buga\inst{1}, Sorana Tania Neme\cb{s}\inst{1}, Atif Mashkoor\inst{2}}

\authorrunning{Addressing Client Needs for Cloud Computing using Formal Foundations}


\institute{Johannes Kepler University of Linz, Austria\\
a.buga$\mid$t.nemes@cdcc.faw.jku.at
\and
Software Competence Center Hagenberg GmbH, Austria\\ 
atif.mashkoor@scch.at
}

%

\toctitle{}
\tocauthor{Andreea Buga and Sorana Tania Nemes}
\maketitle

\begin{abstract}
Cloud-enabled large-scale distributed systems orchestrate resources and services from various providers in order to deliver high-quality software solutions to the end users. The space and structure created by such technological advancements are immense sources of information and impose a high complexity and heterogeneity, which might lead to unexpected failures. In this chapter, we present a model that coordinates the multi-cloud interaction through the specification, validation, and verification of a middleware exploiting monitoring and adaptation processes. The monitoring processes handle collecting meaningful data and assessing the state of components, while the adaptation processes restore the system as dictated by the evolution needs and sudden changes in the operating environment conditions. We employ Abstract State Machines to specify the models and we further make use of the ASMETA framework to simulate and validate them. Desired properties of the system are defined and analysed with the aid of the Computation Tree Logic.
\end{abstract}
\section{Introduction}
\label{sec:intro}

The cloud computing business model offers a wide set of services on a pay-as-you-go payment method, which proves to be efficient for a growing number of businesses. As defined by the National Institute of Standards and Technology (NIST)~\cite{Mell2011}, cloud computing allows ``ubiquitous, convenient, on-demand network access to a shared pool of configurable computing resources''. In most situations, the client needs to trust the provider that the interaction with the cloud is trustworthy and secure. This can lead further to vendor lock-in issues or lack of client control over privacy policies and quality standards.

The research work of the Christian Doppler Client-Centric Cloud Computing Laboratory (CDCC)\footnote{http://cdcc.faw.jku.at/}, headed by Professor Klaus-Dieter Schewe, addresses the lack of client-orientation and the scarcity of formal foundations of cloud services~\cite{Schewe2011b, Schewe2011a}. Its goal is to achieve transparency and reliability of the services, while meeting the client needs. Therefore, various aspects regarding client-cloud interaction have been considered including identity and authorisation management, monitoring of Service Level Agreement (SLA) compliance, detection of malicious security attacks and service content adaptivity to different devices. As a result, several components handling important aspects for the cloud users have been developed as part of the project~\cite{Vleju2015, Rady2015, Lampesberger2016, Holom2016}. All these components have been integrated in a Client-Cloud Interaction Middleware (CCIM), which has been formalized using the Abstract State Machines (ASM) theory~\cite{Borger2003} and ambient calculus~\cite{Cardelli2000}. This middleware will be extended in this chapter with monitoring and adaptation services.

The continuous evolution of computing requirements and the variety of service offerings have led to the need of accessing and composing resources from different providers. A survey focused on cloud buyers and users carried out by RightScale\footnote{https://www.rightscale.com/lp/state-of-the-cloud} in January 2017 reveals that among the companies relying on clouds, eighty five percent of them use now a multi-cloud strategy \cite{RightScale2017}, running services in an average of four clouds (following both public and private deployment models). The initial research work of CDCC has been expanded in order to cover also the requirements of interoperable Cloud-Enabled Large-Scale Distributed Systems (CELDS). The main goal is to enable interaction of services located on different clouds and to ensure their availability and reliability to the end user. The direction given by the work carried out for the CCIM~\cite{bosa:2014} is expanded with a new abstract machine model organized on three layers responsible for the execution, monitoring and adaptation of services running in CELDS.

The current chapter focuses on the description of the aforementioned client-oriented cloud services and their role in fulfilling meaningful aspects for the users like privacy, security and Quality of Service (QoS). The work further investigates the requirements of CELDS with respect to monitoring and adaptation, which aim to increase the availability and reliability. 

The main contribution of this chapter is as follows. First, we propose a fault-tolerant monitoring framework relying on redundancy and partial hierarchy, and discuss a set of metrics to be collected and evaluated. The monitors are considered to be prone to failures and are characterized by a confidence degree value. Second, we define an adaptation layer, which receives information about faulty situations from the monitors. The adaptation solution relies on a case-based repository to build suitable reconfiguration plans in terms of actions and controllers to restore the system to a normal working mode. The workflow of both the monitoring and adaptation frameworks is captured formally in terms of ASMs, whose behaviour is validated and desired properties are model checked with the aid of the ASMETA toolset~\cite{Arcaini2011}\footnote{http://asmeta.sourceforge.net/}.

The remainder of the chapter is organised as follows. In order to have the overview of the project frame, Section~\ref{sec:client-cloud} introduces the client-cloud interaction middleware component and its constituent solutions defined for fulfilling the client needs on single clouds. The description completes the specification of the CCIM, whose interaction leads to the formation of CELDS. The layers of CELDS are defined in Section~\ref{sec:intercloud}. The monitoring and adaptation solutions for CELDS are formally specified in Section~\ref{sec:formal_spec}. The validation and verification procedures for both models are captured in Section~\ref{sec:validation_verification}. Related work is discussed in Section~\ref{sec:related_work} and Section~\ref{sec:conclusion} concludes the chapter.

\section{Overview of the Client-Cloud Interaction System}
\label{sec:client-cloud}
Although clouds aim to meet the service requirements of the clients, they offer users a limited control and transparency in the way their data is handled. In response to this drawback, a client-oriented middleware has been proposed~\cite{Bosa2014}. The middleware encompasses a set of specific services described throughout the current section. We present in this section an overview of the offered services of the middleware, which were previously researched and developed at CDCC under the supervision of Professor Klaus-Dieter Schewe.

\subsection{Client-Cloud Interaction Middleware}
\label{subsec:ccim}
The CCIM defined in \cite{bosa:2014} proposes a high-level formal model of a novel cloud service component based on ambient ASM, which is able to incorporate the major advantages of ASMs and ambient calculus~\cite{Cardelli2000}. The formal models of distributed systems including mobile components are described in two abstraction layers. Therefore, the algorithms of executable components (agents) are specified in terms of ASMs and their communication topology, locality and mobility are described using ambient calculus. The ambient ASM specification gives us a universal way to handle client-cloud interaction independent from particularities of certain cloud services or end-devices, while the instantiation by means of particular ambients results in specifications for particular settings. Thus, the architecture is highly flexible with respect to additional end-devices or cloud services, which would just require the definition of a particular ambient.

\begin{figure}[htb]
	\centering
	\includegraphics[width=0.8\linewidth]{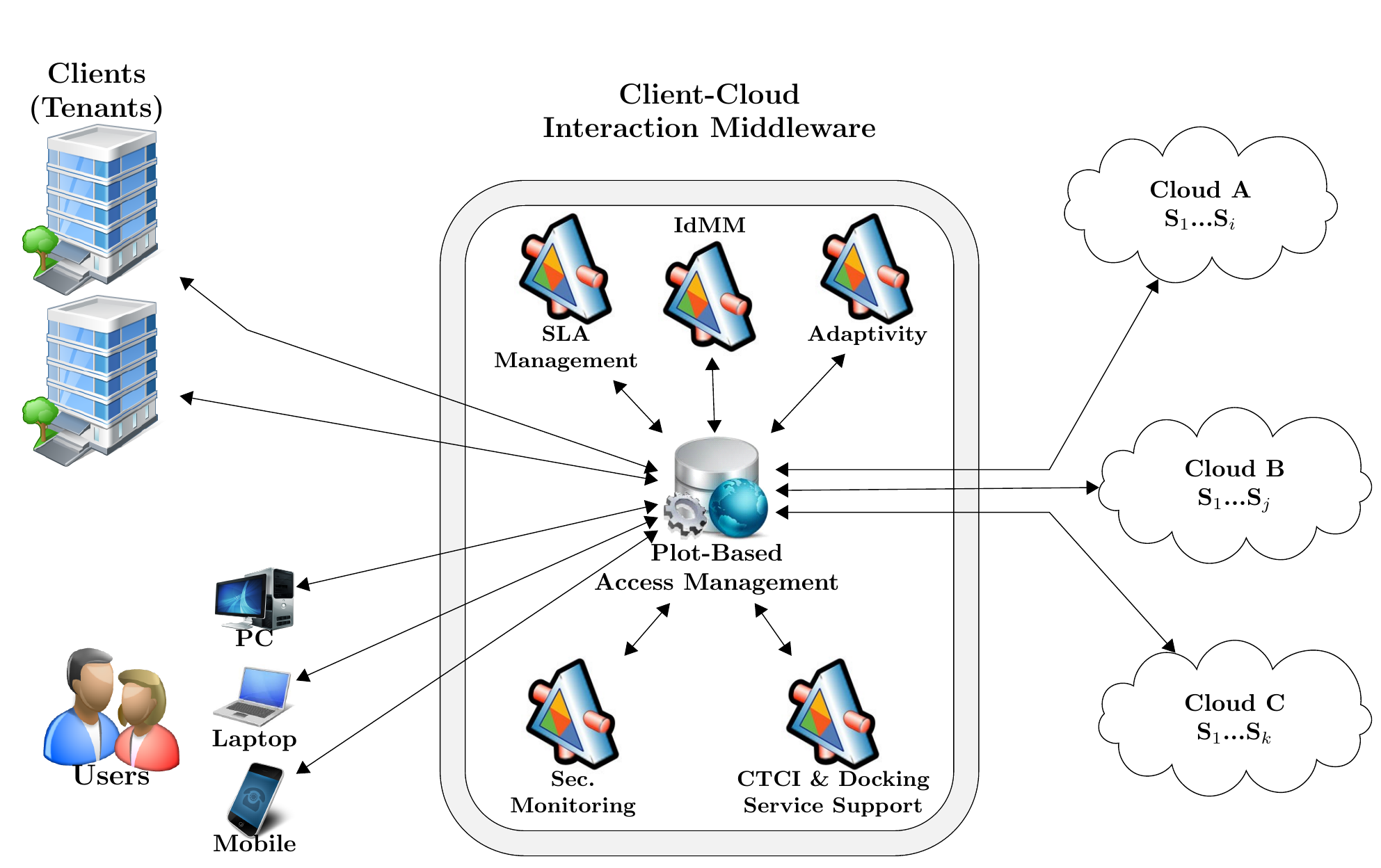}
	\caption{Distributed deployment model for the proposed CCIM, reprinted from~\cite{bosa:2014}}
	\label{fig-ccim}
\end{figure}

This robust architecture model integrates several novel and loosely coupled software solutions (e.g. Client-to-Client Interaction (CTCI) Feature, Identity Management Machine (IdMM), Content Adaptivity, SLA Management, and Security Monitoring Component) into a compound single software component on the client side, see Fig.~\ref{fig-ccim}. 

\subsection{Identity Management Solution for the Cloud}
\label{subsec:idmm}
Although widely spread, cloud computing still presents some vulnerabilities, which make clients reluctant to trust the providers. The report conducted by~\cite{csa:2016} named insufficient identity, credential and access management and also account hijacking as main concerns of users with respect to cloud services. While there are several deployment models of an IdMM surveyed by~\cite{sturrus:2011, zwattendorfer:2014}, cloud providers prefer the approach that keeps on their servers all the data needed for authentication and authorization. 

The CCIM offers an IdMM service that can deal with three interaction scenarios~\cite{Vleju2015}. First, the IdMM allows the user to save credentials directly using a cloud service, which shifts the responsibility to the cloud, but might also affect data privacy. Second method permits the user to obfuscate part of the information. A third approach uses federated identity management protocols. This offering of the middleware allows the client to decide on the level of control given to the cloud for storing sensitive information. 

\subsection{Monitoring of Client-Cloud Interaction using Service Level Agreements}
\label{subsec:sla}

Each cloud provider offers a set of quality measures of its services, which comes in the form of an SLA contract. SLAs stipulate that whenever described properties are not fulfilled, the provider must pay a compensation or a penalty to the client~\cite{milanovic:2011}. For this reason, a fair SLA management platform, which ensures a transparent communication between the client and the provider is needed. The middleware component has been enriched, in this sense, with a framework modelled using a Web Ontology Language (OWL) which allows automated reasoning and monitoring of SLAs. The client is also granted the right to edit, accept or reject an SLA contract~\cite{Rady2015}. Granting both sides permission to modify the SLA document requires a synchronization policy in order to avoid possible inconsistencies.~\cite{Rady2015} proposes an approach based on Lamport's synchronization algorithm, which uses logical clocks~\cite{Lamport1978}.


\subsection{Anomaly-Based Intrusion Detection}
\label{subsec:intrusion}
As cloud computing accumulates data on the provider side, clouds and their clients become vulnerable to security attacks like malicious insiders, tampering, data breaches, or loss of data. A deep understanding of the problem that enables attacks is necessary to develop countermeasures and secure software.

~\cite{Lampesberger2016} discusses vulnerabilities in terms of formal language theory and proposes a protocol monitor that mitigates vulnerabilities by validating in- and outgoing messages of a web component, i.e. service provider or consumer. The research proposes a language-based learning approach for anomaly detection on tree-based documents like eXtensible Markup Language (XML) using pushdown automaton. Datatypes XML visibly pushdown automaton (dXVPAs) is inferred to capture the expected language at a particular service interface and is translated for stream validation. 

\subsection{Adaptivity of Cloud Content}
\label{subsec:cloud_adaptivity}

In the context of cloud services, a well-founded system should not tailor specific applications to each type of end-device (desktop computers, laptops, tablets, smartphones, etc.), but guarantee that its clients are able to access the same cloud service and its subsequent output regardless of the used end-device, its operating system and/or distinct hardware characteristics (e.g. processor speed, display size and resolution).

To tackle the problem of providing cloud services to different devices, an adaptivity component is created as part of the middleware application~\cite{Holom2016}.~\cite{Holom2016a} proposes making use of various internal components to manage the interaction between the clients and the cloud services, so that on-the-fly layout and content adaptation (mostly needed in case of mobile devices) are ensured. The adaptivity component envisions a web application as a communication channel which automatically detects device properties on the client-side and uses them on the server-side to modify the content coming from the cloud. 


\section{A Management System for Interconnecting Clouds}
\label{sec:intercloud}
Interconnected clouds were defined as a model, which permits reallocation of resources and workload transfer through the collaboration of services of different cloud providers in order to achieve the promised service quality measured in terms of performance, availability and agreed SLA~\cite{aoyama2011inter}.

This section introduces the concept of a client-oriented distributed middleware for multi-cloud services and highlights the structure and role of the monitoring and adaptation layers. Multi-cloud systems are already faced with numerous challenges (e.g. complexity, network failures, and bottlenecks) and the addition of other components like monitors and adapters should not interfere with the execution of processes. In order to better understand the intrinsic problems that the framework can face, we focused our attention on an ASM formal model, which can complement the existing work in the area. 
\subsection{Overview of the Distributed Cloud Interaction Middleware}
\label{subsec:intercloud_middleware}
\begin{figure}[htb]
	\centering
	\includegraphics[width=1\linewidth]{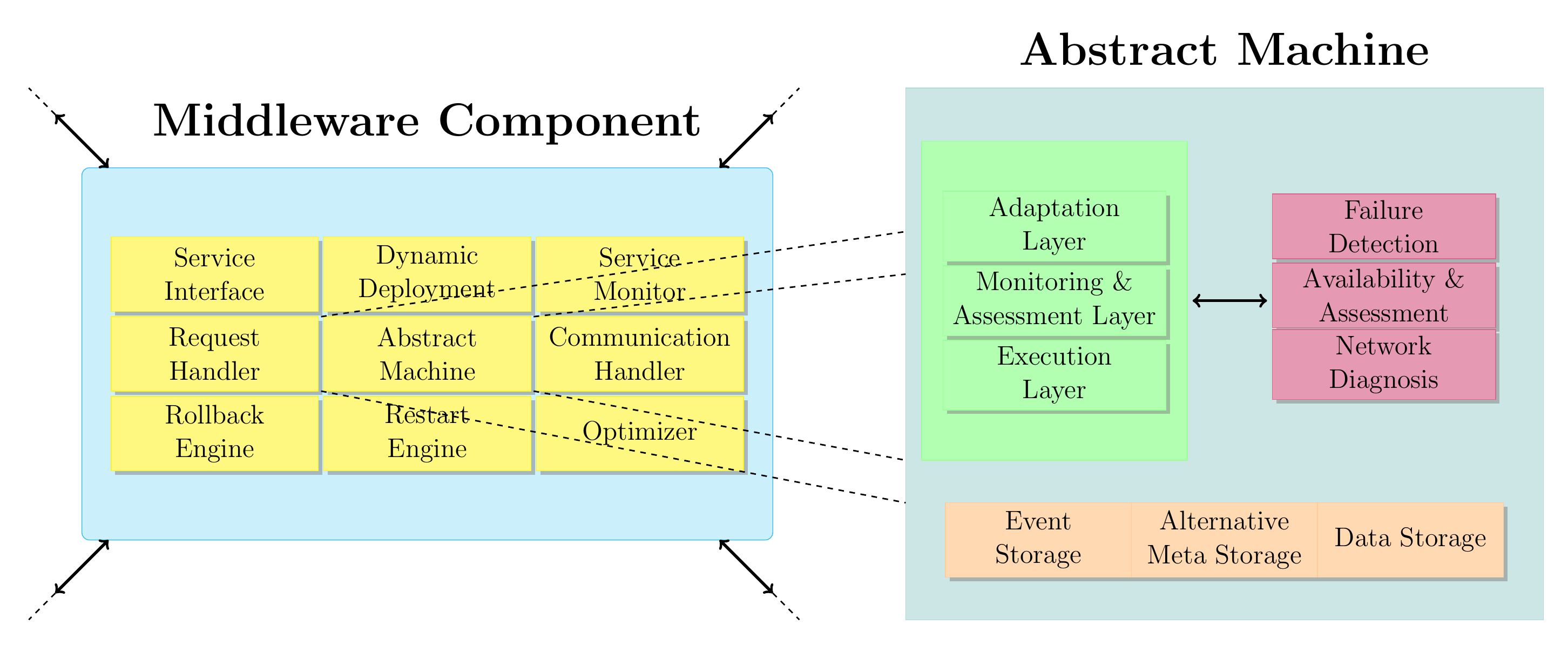}
	\caption{Layered structure of a middleware component, reprinted from~\cite{buga2017a}}
	\label{fig:ground}
\end{figure}

As illustrated in Fig.~\ref{fig:ground}, the middleware consists of several machines and engines used for supporting the three layers (execution, monitoring and adaptation) of the abstract machine. The execution layer establishes the communication between nodes, handles requests from the users and permits the description of service interfaces \cite{bosa:2014}. The monitoring layer performs a continuous assessment of the status of the nodes, with a focus on detecting unavailability and crash failures, and communicates information about such issues to the adapters. The adaptation layer analyses a reported issue with previous solved cases and proposes a reconfiguration plan. The efficiency of the adaptation processes is then evaluated by the monitors.

The collaboration of the monitoring and adaptation layers aims to improve the reliability of the system and to optimize the reconfiguration processes. The three components of the storage refer to the information locally logged in the system. The data part includes low- and high-level metrics collected by the monitors, while adaptation processes are saved in the event storage. Supplementary information (characteristics of the node, uptime of monitors and adapters) are stored in the alternative meta storage. The structure and workflow of monitoring and adaptation processes are extended with a formal specification in Section~\ref{sec:formal_spec}.


\subsection{Monitoring Layer}
\label{subsec:monitoring_layer}
Monitoring refers traditionally to collecting specific data from system components and evaluating them in order to discover execution problems, availability and performance issues. Monitors are components of CELDS and they are also prone to failures. While there is a broad spectrum of issues to monitor, we focus on availability, latency and crash failures. Instead of addressing a wider set of problems, we rather emphasize the accuracy of monitoring processes. Our proposed approach aims to mitigate the risk of monitor failures through enforcing redundancy and applying a confidence function that would correspond to the quality of the diagnoses established by a monitor.

The monitoring layer of the middleware handles the collection of relevant information from the execution. Monitors assess the status of each node of the system. Being part of the CELDS, monitors also face possible failures. In order to avoid the problem of a single point of failure, we opted for assigning a set of monitors to each node of the system and to evaluate possible issues with the aid of a collaborative decision. Also, each monitor is assigned a confidence measure, which reflects its performance. Monitors with low values are stopped by the middleware agent and either restarted or replaced.

We consider that a middleware component handles the assignment of monitors to a node. The middleware is also responsible for electing a leader for each set of monitors. The leader is in charge of acquiring monitoring data from each monitor and establishing a common diagnosis whenever an issue is reported. 
%


\subsubsection{Architecture of the System}
\label{subsubsec:mon_architecture}
\begin{figure}
\centering
\includegraphics[width=0.8\textwidth]{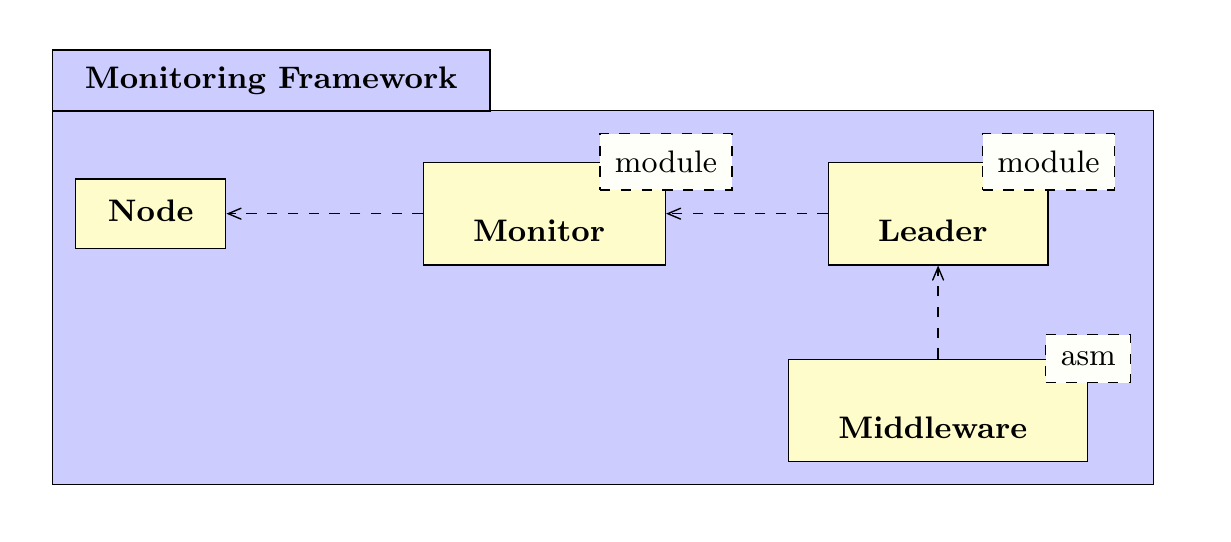}
\caption{Structure of the monitoring framework, adapted from~\cite{buga2017b}}
\label{fig:structure}
\end{figure}

The structure of the monitoring framework depicted in Fig.~\ref{fig:structure} reflects the dependencies between its components. We opted for a solution composed of different modules and an ASM machine. We assigned the main rule to the middleware, which calls also the execution of the modules. We propose ensuring reliability through redundancy of the monitors. Therefore, a node is evaluated by a set of monitors, which are coordinated by a leader. The leader introduces a hierarchical view to the approach and is responsible for coordinating information obtained from different monitors. Leader module relies on the monitor module, requiring it the necessary information for assessing a collaborative diagnosis. The monitor contains a reference to the node it is assigned to.
%

\subsection{Adaptation Layer}
\label{subsec:adaptation_layer}
The adaptation layer reacts to the results of the monitors and identifies changes to CELDS by replacing (sets of) services, relocation of services, deployment and integration of new services, replication, removal of replication, etc. The guideline for the selection of the best alternative is the repair of the encountered problem under presumably optimal performance. Based on the decision which alternative implementation of the CELDS is to be taken, the running system will be (partially) interrupted, rolled back to a consistent state and restarted with the new remaining execution plan.

Regardless of the delivery model associated with the services being provided and consumed in the cloud environment (Software as a Service (SaaS), Platform as a Service (PaaS), Infrastructure as a Service (IaaS), there will be entities and relationships in different layers of the cloud environment that can request or cause adaptations to each other, which in return may lead to conflicting actions. To this purpose, the Adaptation Engine must be general enough to be able to automatically coordinate among the different layers, entities and adaptation options while perpetually reacting to the notifications from the monitoring component. Therefore, once the system recognizes a violation of the expected behaviour, it enters a reconfiguration state and triggers a self-adaptive process in order to reach a stable and correct behaviour.

\subsubsection{Architecture of the System}
\label{subsubsec:adap_architecture}
As an inner component of the abstract machine included in the middleware, the Adaptation Engine provides a unified solution using a Case-Based Reasoning approach (CBR)~\cite{Aamodt1994} enhanced in matters of adaptation actions, their usage and impact on the system. There are two major parts to be considered for the adaptation process: the decision phase defined by solution exploration, identification and maintenance, and the solution management and enactment phase, each with well delimited responsibilities and areas of inference and control. 

In the envisioned framework, a case for system adaptation \textit{C\textsubscript{r}} represents a formatted instance of a problem \textit{P\textsubscript{r}} linked to a recorded solving experience \textit{S\textsubscript{r}}. The problem part \textit{P\textsubscript{r}} represents a collection of description features (e.g. response time, price, portability, region, availability, input/output bandwidth) which, as described in~\cite{Nemes2017}, are subject to similarity functions and common pattern recognition mechanisms. The adaptation solution is selected based on the collected knowledge of failure scenarios and associated previously validated solutions. The solution part \textit{S\textsubscript{r}} is configured and stored in the repository as a workflow schema detailing the actions and underlying transition dependencies needed to restore the system to a normal execution mode.

An action is an autonomous entity (e.g. a software module) which has the power to act or cause a single update to the system that helps to solve the problem the adaptation system was employed for. Its autonomy and self-awareness imply that its execution is not controlled by the environment or other actions but is, thus, able to deal with unpredictable, dynamically changing, heterogeneous environments while relying fully on existing solutions for CELDS adaptability. Examples of actions for system adaptation include discovery of a suitable matching service to replace the problematic one (by accessing the capabilities of an existing tool) or the dynamical reconfiguration of service calls to the new service. The adaptation is thus a trace of actions that are being executed by employing their capabilities to resources to fulfil a certain adaptation that would bring the system to a normal execution state.

The action's instantiation and execution are handled by linked ActionController loaded based on the defined contract for that particular action. The actions' ordering and dependency on other actions is handled by means of notification/signalling, where every action state change triggers a signal broadcast raised by the parent ActionController. The responsibility of the ActionController lies in the observation of its states, the subsequent action and its states and data, and the detection of undesirable behaviour when executing the action or communicating with other action controllers. 

\begin{figure}
  \centering
\includegraphics[width=0.8\textwidth]{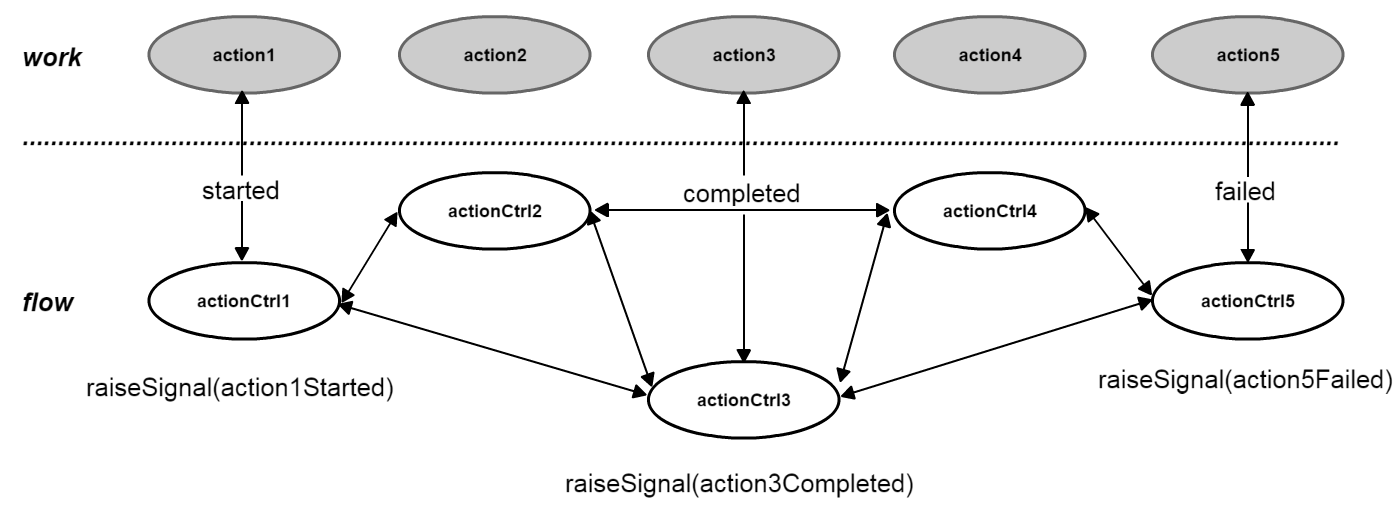}
\caption{Overview of workflow schema and execution.}
\label{fig:workflow-overview}
\end{figure}

Therefore, the adaptation system consists of a finite set of interacting ActionControllers. Based on the assessed observations and broadcast notifications triggered by actions' execution or failure, the ActionControllers affect the system towards the remediation of the reported problem/failure. The assessment implies either enacting and executing its corresponding action, or ignoring the notification as it is not of interest in the given solution configuration. This model's underlining observer/controller architecture is one realisation of the feedback loop principle, and guides the system behaviour and dynamics for the adaptation to succeed in reaching the intended goals. Fig.~\ref{fig:workflow-overview} depicts the overall structure of the adaptation process once the problem is mapped to previous encountered problems and the attached solution is carried out based on its configuration. 

Having ActionControllers to monitor and handle the interaction between the actions of a solution emphasizes new properties of the actions being defined in terms of needed input, concrete implementation and resulting output. In addition, it enables the possibility to easily add or substitute any given number of adaptation actions. Once an action is altered or added, the workflow schema needs to reflect the required changes and dependencies with the other configured adaptation actions. These changes are later on translated and visible in the ActionControllers and the interaction between them.


\section{Formal Specification of the Monitoring and Adaptation Processes}
\label{sec:formal_spec}

\subsection{Background on the ASM Formal Method}
\label{subsec:asm_background}
Throughout time, several formal methods have been proposed, each dealing with different properties to model. In order to better understand the suitability of the methods for CELDS we consulted the survey elaborated in~\cite{Kossak2016}. The survey verifies a set of aspects (e.g. supported development phases, tool support, modelling capabilities, and industrial applicability) for each method.

The two most-suitable candidates for modelling a distributed system and its specific characteristics like concurrency and non-determinism are, according to the survey, ASMs and TLA+~\cite{Lamport2009}. We finally opted for adopting ASMs for a couple of other properties like the assistance throughout the software development process and their suitability for industrial applications. Other important candidates for specifying our proposed monitoring and adaptation services are Petri Nets and the Unified Modelling Language (UML). However, UML still lacks a formal precision and Petri Nets further proved to be more verbose for a set of specific distributed systems applications~\cite{Borger2016}. 

ASMs extend the concept of Finite State Machine (FSM) with the possibility to specify synchronous parallel operations and enhance the input and output states with data structures. An ASM machine consists of a tuple $M = (\Sigma, S_0, R, R_0)$, where $\Sigma$ is the signature (the set of all functions), $S_0$ is the set of initial states of $\Sigma$, $R$ is the set of rule declarations, and $R_0$ is the main rule of the machine. Rules of an ASM follow the \textbf{if} \textit{Condition} \textbf{then} \textit{Updates} structure~\cite{Borger2003}, where the \textit{Updates} can consist of a set of parallel assignments of values to locations.

ASMs are able to capture different functions and their structure highlights the separation of concerns principle. Static functions refer specifically to constants, while dynamic functions can be updated during execution. \textit{Controlled} functions are written only by the machine, while \textit{monitored} ones are written by the environment and read by the machine. Both the machine and its environment can update \textit{shared} functions.

The CoreASM~\cite{Farahbod2007} execution engine and the ASMETA toolset support the definition of ASM models through their own specific ASM dialects. Each of them permits different functionalities to simulate, check or visualize ASM models. An attempt to unify both dialects led to the development of the Unified Syntax for Abstract State Machine~\cite{Arcaini2016}, which supports also the translation of models to C++ code designed for Arduino devices~\cite{Bonfanti2017}. 

Although the ASM method permits a good formalization it still has a set of drawbacks as follows. Time related aspects, which are critical for real systems cannot be integrated in the model. On a very abstract level, one can constraint a loop block of rules to execute for at most a certain number of times, without actually integrating a time measurement. For validation, the usage of composed functions is limited. In the verification stage, infinite domains have to be removed or replaced by finite sets or enumerations. However, the models are significant and provide important insights on the processes workflow.
\subsection{Specification of the Monitoring Solution}
\label{subsec:monitoring_spec}

As presented in Section \ref{subsubsec:mon_architecture}, the formal model of the monitoring layer relies on two modules and an ASM component. The node is defined simply as a domain inside the monitor module and no rules and functions for it are expressed. The following subsections address specific aspects related to each of the components and capture parts of the ASM model\footnote{A more detailed specification can be found at \url{http://cdcc.faw.jku.at/staff/abuga/esocc.zip}}. The proposed specifications were built entirely in one step, but another approach would be to incrementally construct the models through refinements as described in~\cite{Borger2003}.
\subsubsection{Monitor Module}
\label{subsubsec:monitor}
The monitor module represents the core of the solution. After being assigned to a node, the monitor is in the \textit{Active} state from which it sends a heartbeat request to verify the availability of the node and its latency, and moves to the \textit{Wait for response state}. In this state, the monitor verifies two guards. First, it verifies whether a response to its request is received. If so, it checks whether the delay of the response is acceptable. If this condition is satisfied, the monitor moves to the \textit{Collect data} state. If the request has no response or the delay is too big, the monitor moves to the \textit{Report problem} state. In the \textit{Collect data} state, data about the status of the node is gathered (CPU, memory, storage). The monitor attempts then to \textit{Retrieve data} from a local storage containing previous logs. If the repository is available, the monitor queries it. 
The monitor moves afterwards to the \textit{Assign diagnosis} state, where it interprets all the available data. In case problems are discovered, the monitor moves to the \textit{Report problem} state, otherwise it transits to the \textit{Log data} state, where meaningful data and operation are locally saved. When an issue is discovered, the monitor modifies a constraint that triggers a request towards the leader of the node to start a collaborative diagnosis. After reporting the issue, the monitor moves to the \textit{Log data} state. The ground model of the module has been detailed previously also in~\cite{buga:er2017}.
\subsubsection{Leader Module}
\label{subsubsec:leader}
The leader module contains rules for collecting assessments from each of the monitors and building a collective diagnosis. In the current version of the model, we consider that the information submitted by each monitor is equally important for the collective result as shown in Code~\ref{lst:request_data}. 
\begin{lstlisting}[frame=single, language=AsmetaL, basicstyle=\scriptsize, caption={Request data rule of the leader module}, label={lst:request_data}]
	rule r_RequestData ($l in Leader) = 
		choose $node in Node with (eq(has_leader($node), $l)) do
			forall $m in Monitor with (assigned_node ($m) = $node) do
				if (isDef(diagnosis ($m))) then
					if (diagnosis ($m) = NORMAL) then
						normal_diagnoses ($l) := normal_diagnoses ($l) + 1
					else if (diagnosis ($m) = FAILED) then 
							failed_diagnoses ($l) := failed_diagnoses ($l) + 1
						else
							critical_diagnoses ($l) := critical_diagnoses ($l) + 1
						endif
					endif
				endif
\end{lstlisting}
Assessing the status of a node by the leader implies choosing the diagnosis proposed by the majority of the monitors assigned to the node. Code~\ref{lst:assess_node} contains the responsible rule. At the end of a collaborative diagnosis round, each of the monitors updates its confidence degree according to a function defined by~\cite{buga2017c}. This function takes into account the similarity of the information submitted by a monitor in comparison with the one provided by other monitors assigned to the same node. Therefore, the lower the similarity, the higher the penalty. The rule for recalculating the confidence degree is left abstract for this version of the model. After recalculating the confidence and saving the diagnosis, the counters for failed, critical and normal diagnoses are reset.
\begin{lstlisting}[frame=single, language=AsmetaL, basicstyle=\scriptsize, caption={Assess node rule of the leader module}, label={lst:assess_node}]
	rule r_AssessNode ($l in Leader) = 		
		if (max(failed_diagnosis ($l), critical_diagnosis ($l)) = failed_diagnosis ($l) ) then
			if (max(failed_diagnosis ($l), normal_diagnosis ($l)) = failed_diagnosis ($l) ) then
				assessment ($l) := FAILED
			else
				assessment ($l) := NORMAL
			endif
		else
			if (max(critical_diagnosis ($l), normal_diagnosis ($l)) = critical_diagnosis ($l) ) then
				assessment ($l) := CRITICAL
			else
				assessment ($l) := NORMAL
			endif
		endif
\end{lstlisting}
\subsubsection{Middleware ASM}
\label{subsubsec:middleware}
The middleware component orchestrates the processes of the monitor and leader modules and manages the workflow of the system. It initializes the functions and contains the main rule, which calls for the execution of rules belonging to the monitor and leader modules. The middleware assigns a set of monitors to each node and elects a leader (these processes occur only once as in each of the corresponding rules there is a guard verifying whether monitors were already assigned to a node and whether a leader has already been elected). The middleware ensures that when a monitor wants to report a problem (trigger\_gossip), the corresponding leader moves to the \textit{Evaluate} state and starts to collect diagnoses from all the monitors assigned to the node reported as having problems. The middleware also dismisses monitors whose confidence degree is below the accepted minimum value.
\begin{lstlisting}[frame=single, language=AsmetaL, basicstyle=\scriptsize, caption={Main rule of the middleware agent}, label={lst:middleware}]
	rule r_MiddlewareProgram = 
		if (middleware_state(self) = EXECUTING) then
			par
				r_AssignMonitorsToNode []
				r_ElectLeader []
				forall $m in Monitor with (trigger_gossip($m)) do
					let ($n = assigned_node($m)) in
						let ($l = has_leader($n)) in
							if (leader_state($l) = IDLE_LEADER) then
								leader_state($l) := EVALUATE
							endif
						endlet
					endlet			
				forall $mon in Monitor with (confidence_degree($mon) < min_confidence_degree) do
					r_DismissMonitor [$mon]
			endpar
		endif	
\end{lstlisting}

\subsection{Specification of the Adaptation Solution}
\label{subsec:adaptation_spec}
Based on the overall specification of the adaptation framework described in Section \ref{subsec:adaptation_layer}, we define the specific states and transitions of the adaptation processes, with emphasis on one of the main modules, the actionController module.

\subsubsection{ActionController Module}
\label{subsubsec:actionController_module}
The actionController module represents the core of the action enactment process of the system and bares the responsibilities that come with the action's observation and triggering, compliant to the executed workflow schema. The actionController can pass through several states by various rules and guards.

\begin{lstlisting}[frame=single, language=AsmetaL, basicstyle=\scriptsize, caption={Acknowledge notification ASM rule.}, label={lst:acknowledgeNotification}]	
rule r_AcknowledgeNotificationReceived($c in Controller,$broadcaster in Controller) =
	if (controller_state($c) = NOTIFICATION_RECEIVED) then
		seq
			controller_state($c) := ASSESS_NOTIFICATION
			par
				acknowledged_controllers($broadcaster) := acknowledged_controllers($broadcaster) + 1
				r_HandleNotification[$c]
			endpar
		endseq
	endif  
\end{lstlisting}

At initialization, the actionController is in the \textit{Passive/Waiting notification} state. Once a notification arises, the actionController acknowledges the received notification in disregard of the actual sender, after which it moves to the \textit{Assess notification} state. The rule responsible for acknowledging a notification is captured in Code~\ref{lst:acknowledgeNotification}.

In order to assess the received notification, the actionController must first validate the received notification. If the notification is compliant with its contract, the actionController broadcasts the notification/signal that the underlying action is bound to start its execution, as captured in Code~\ref{lst:broadcastNotification}. 
\begin{lstlisting}[frame=single, language=AsmetaL, basicstyle=\scriptsize, caption={Broadcast notification ASM rule.}, label={lst:broadcastNotification}]
rule r_BroadcastNotification($c in Controller, $n in Notification) =
	forall ($neighbor in Controller) then
		if (not(id($c) = id($neighbor)) 
			seq
				acknowledged_controllers($c) := 1					
				par						
					controller_state($c) := WAITING_FOR_ACKNOWLEDGEMENT
					AcknowledgeNotificationReceived[$neighbor, $c]
				endpar
			endseq
		endif
	endforall
\end{lstlisting}

Every actionController instantiated as part of the same adaptation session must receive and acknowledge the broadcast notification. Once the notification is acknowledged by all neighbouring actionControllers, the actionController should proceed to the execution of its action. The rule responsible for triggering the associated adaptation action is captured partially in Code~\ref{lst:triggerAction}.  

\begin{lstlisting}[frame=single, language=AsmetaL, basicstyle=\scriptsize, caption={Trigger action ASM rule}, label={lst:triggerAction}]
	rule r_TriggerAction($c in Controller) =
		seq
			while (controller_state($c) = RUNNING_ACTION)
				wait
			if (action_completed($c))
				par
					r_BroadcastNotification[$c, ACTION_COMPLETED]
					r_AwaitAcknowledgement[$c]
					if (acknowledged_controllers($n) = numberOfControllers)
						par						
							r_ClearNotificationEcho[$c]	
							controller_state($c) := WAITING_NOTIFICATION
						endpar
					else
						par
							controller_state($c) := CONTROLLER_ACKNOW_FAILED
							AssessDataAndStatus
						endpar
					endif
				endpar
			else
				par
					BroadcastNotification[$c, ACTION_FAILED]
					...
\end{lstlisting}

Regardless of the output of the executed action, one notification will be broadcast signalling the success or failure of this particular system update. There is no linked track of the actionControllers' order to execution. Therefore, if at least one action fails or one actionController does not acknowledge any of the sent notifications, the adaptation is abruptly terminated. The component data and status are afterwards assessed and logged accordingly. If the associated action's successful execution and acknowledgement by all the other controllers are fulfilled, the actionController reaches again the initial state. This initial state is reached again either when the associated action's execution and acknowledgement by all the other controllers are fulfilled or when the received notification is not bound to influence the actionController in question. This is a clear indication of the continuous character of the adaptation process which takes place in the background of service execution.

\section{Validation and Verification of the Specifications}
\label{sec:validation_verification}
This section presents the practicability of the ASM method in reasoning about the system requirements by applying different validation and verification activities. In Section~\ref{subsec:validation}, the usefulness of scenario-based validation is explained and examples of such simulated scenarios are provided. Section~\ref{subsec:verification} explains how model checking techniques are applied using the AsmetaSMV tool and presents the verification of some classical temporal properties specified on the models of the adaptation and monitoring solutions.
\subsection{Validation}
\label{subsec:validation}
Validating the models enables us to check whether the system behaves as expected and the models correctly capture the intended requirements. We performed random and interactive simulation with the aid of AsmetaS simulator~\cite{Gargantini2008a}. The AsmetaV validator allowed us to also build and execute scenarios of expected behaviours. A scenario is usually associated with a specific execution path and can be used for testing the state of the system after a set of transitions.

Inconsistency errors were detected during simulation with the aid of the AsmetaS tool. For example, more than one system failure can be reported in a short time frame. And although a schema is locked while its associated solution is executed, a parallel execution of simultaneous adaptations may try to update system parts or components with different values at the same time. Triggering simultaneously multiple adaptations within the system is then supported by transaction specific operations where every solution is annotated with extensive knowledge on the area of inference in the system of each subsequent action, which would later on be considered in the adaptation decision phase.

Model simulation leads sometimes to verbose traces, from which it is hard to discover possible design issues. For such situations, we opted for the use of validation scenarios defined with the aid of the Avalla language introduced in~\cite{Carioni2008}. The scenarios allow the description of particular execution paths and through their validation we can check whether boolean constraints are fulfilled. We present in Code~\ref{lst:scenario} a simple scenario for determining whether the leader correctly evaluates the diagnosis of a node observed by three monitors. The first monitor indicates a high latency, and thus triggers the leader to evaluate the status of the node. At the time of the request, the second and third monitor did not carry out a full monitoring cycle and they cannot send any evaluation. For this situation, the leader takes into consideration only the available information from the first monitor and evaluates correctly that the node has failed. This situation, however, highlights the fact that insufficient data might lead to inconsistent evaluations.
\begin{lstlisting}[frame=single, language=AsmetaL, basicstyle=\scriptsize, caption={AsmetaV simple scenario for leader diagnosis validation}, breaklines=true, label={lst:scenario}]
set assigned_monitors(node_1) := [monitor_1, monitor_2, monitor_3];
set has_leader(node_1) := leader_1;
set leader_state(leader_1) := IDLE_LEADER;
set has_leader(node_1) := leader_1;
step
set heartbeat_response_arrived(heartbeat_1) := false;
set heartbeat_response_arrived(heartbeat_2) := true;
set heartbeat_response_arrived(heartbeat_3) := true;
set heartbeat_latency(heartbeat_2) := 5;
set heartbeat_latency(heartbeat_3) := 7;
step
set heartbeat_response_arrived(heartbeat_1) := true;
set heartbeat_latency(heartbeat_1) := 21;
set monitor_measurements(monitor_2) := [("Latency", 5), ("CPU Usage", 10), ("Storage Usage", 15), ("Memory Usage", 10), ("Bandwidth", 50)];
set monitor_measurements(monitor_3) := [("Latency", 7), ("CPU Usage", 40), ("Storage Usage", 15), ("Memory Usage", 10), ("Bandwidth", 30)];
step
set is_repository_available(monitor_2) := true;
set is_repository_available(monitor_3) := false;
step
step
step
check assessment(leader_1) = FAILED;
\end{lstlisting}

\subsection{Verification}
\label{subsec:verification}
Correctness and reliability of the monitoring and adaptation solutions were guaranteed by verifying classical temporal properties (e.g. reachability, safety, correctness). Through safety properties, we ensure that something bad will not occur. Reachability properties verify whether a certain state can be reached from the initial state of the system. These together with correctness and component specific properties are expressed with Computation Tree Logic (CTL) formulas and are verified with the aid of the AsmetaSMV Eclipse plugin~\cite{Arcaini2010b}, which also relies on the NuSMV model checker~\cite{Cimatti2000}. NuSMV imposes a set of constraints on the model as it works only with finite sets and domains. Hence, we had to simplify the initial models in order to verify them. Several properties targeting different modules of the monitoring and adaptation solutions are listed below. This list of properties extends the one previously defined in~\cite{buga2017b}.\\

\noindent \textbf{Monitor safety property.}
First property, showcased in Code~\ref{lst:prop_1}, indicates that a monitor which waits for a reply for its heartbeat request and which receives a response whose delay did not exceed the maximum limit, moves immediately to the \textit{Collect data} state. This also ensures that a monitor cannot falsely report a problem.
\begin{lstlisting}[frame=single, language=AsmetaL, basicstyle=\scriptsize, caption={Monitor reachability property}, label={lst:prop_1}]
CTLSPEC (forall $m in Monitor with ag((monitor_state($m) = WAIT_FOR_RESPONSE and 
heartbeat_response_arrived($m) and not(heartbeat_timeout($m)))
implies ax(monitor_state($m) = COLLECT_DATA)))	
\end{lstlisting}
\textbf{Leader reachability property.}
Addressing the collaboration between a monitor and a leader, we verify that for all the situations in which a monitor wants to report a problem, the leader eventually moves to the \textit{Evaluate} state. We used a static function for a leader in this case as we could not apply a composed function to point towards the leader state of a leader assigned to a monitor inside the CTL formula.
\begin{lstlisting}[frame=single, language=AsmetaL, basicstyle=\scriptsize, caption={Leader reachability property}, label={lst:prop_2}]
CTLSPEC (forall $m in Monitor with ag((trigger_gossip($m) = true) 
implies ef(leader_state(leader_1) = EVALUATE )))		
\end{lstlisting}
\textbf{ActionController reachability property.}
Considering the collaboration between an ActionController and its subsequent action, we verify that for all the situations in which an action wants to start its execution, the ActionController eventually moves from the state \textit{WaitingForAcknowledgement} to \textit{ActionRunning}.
\begin{lstlisting}[frame=single, language=AsmetaL, basicstyle=\scriptsize, caption={ActionController reachability property}, label={lst:prop_3}]
CTLSPEC (forall $a in Action with ag((trigger execute($a) = true) 
implies ef (actionController state(actionController 1) = ACTION_RUNNING )))
\end{lstlisting}
\textbf{ActionController liveness property.}
Any ActionController which is assigned to an action, being thus in the \textit{ActionRunning} state, eventually reaches the state where it is ready to be removed, after the adaptation process is successfully completed. We ensure in this case that an adaptation cycle is eventually completed.
\begin{lstlisting}[frame=single, language=AsmetaL, basicstyle=\scriptsize, caption={ActionController fairness property}, breaklines = true, label={lst:prop_4}]
CTLSPEC (forall $ac in ActionController with ag((actionController state($ac) = ACTION_RUNNING) implies ef(actionController state($ac)=READY_FOR_REMOVAL)))
\end{lstlisting}
\textbf{Monitor safety properties.} 
The following property verifies that any identified issue (e.g. a high usage value for CPU and memory) is reported to the leader, which triggers afterwards a collaborative diagnosis. If the monitor does not identify an issue, it must not reach the \textit{ReportProblem} state. By checking both these properties, the monitors are prevented to report false positives.
\begin{lstlisting}[frame=single, language=AsmetaL, basicstyle=\scriptsize, breaklines = true, caption={Monitor safety properties}, label={lst:prop_5}]
CTLSPEC (forall $m in Monitor with ag( (is_problem_discovered($m)) implies ax(monitor_state($m) = REPORT_PROBLEM))) CTLSPEC (forall $m in Monitor with ag( monitor_state($m) = ASSIGN_DIAGNOSIS and (not(is_problem_discovered($m))) implies ex(monitor_state($m) = LOG_DATA)))
\end{lstlisting}
\textbf{Leader fairness property.}
This property of a leader deals with reaching a correct assessment. It verifies that the leader agent starts and resets its counters for a diagnosis to zero at the beginning and end of each voting cycle. Therefore, before inquiring the monitors, the leader is not biased towards a specific diagnosis, but it rather starts from the premise there is an equal chance that the observed node is in a normal, critical or failed situation. 
\begin{lstlisting}[frame=single, language=AsmetaL, basicstyle=\scriptsize, caption={Leader fairness property}, label={lst:prop_6}, breaklines = true]
CTLSPEC (forall $l in Leader with ag( (leader_state($l) = IDLE_LEADER) implies ax(failed_diagnoses($l) = 0 and critical_diagnoses($l) = 0 and normal_diagnoses($l) = 0) ))	
\end{lstlisting}
\textbf{Leader property.}
The leader collects information from all the monitors assigned to the node it is responsible for. As each monitor submits only one analysis value, the number of diagnoses the leader gathers in the assessment is equal to the number of monitors assigned to the node. The following CTL property verifies this assumption and ensures that the leader collects the exact number of diagnoses from the monitors. 
\begin{lstlisting}[frame=single, language=AsmetaL, basicstyle=\scriptsize, caption={Leader property}, label={lst:prop_7}, breaklines=true]
CTLSPEC (forall $l in Leader with ag( (leader_state($l) = ASSESS) implies ax(failed_diagnoses($l) + critical_diagnoses($l) + normal_diagnoses($l) = 3) ))
\end{lstlisting}
Code~\ref{lst:prop_ver} displays the positive results for the evaluation of some of the monitor properties (Code~\ref{lst:prop_1}, Code~\ref{lst:prop_2}, Code~\ref{lst:prop_5}, Code~\ref{lst:prop_6}, Code~\ref{lst:prop_7}) tested for a system consisting of a node with three assigned monitors.

\section{Related Work}
\label{sec:related_work}
Previous work in the area of CELDS services focuses on providing robust infrastructures, which enable easy resource and service transition among different cloud providers~\cite{modaclouds2016, Rochwerger2011, Petcu2013}. A higher-level approach has been previously addressed through the development of the TOSCA language, which allows a standardized description of cloud services~\cite{Binz2014}. Our project extends the existent work and addresses the need of formalizing cloud-specific processes. While previous work focuses on service execution, our approach highlights the importance of correct processes.

While the adoption of formal methods in real systems is highly arguable due to the steep learning curve and large time project setting, they can indicate serious problems of the system that might reside from either the analysis or design phase. A relevant work to adopt formal specifications was carried out by Amazon Web Services, which integrated the TLA+ formal language to verify its infrastructure and discover design flaws that cause bugs~\cite{Newcombe2015}. The ASM method we adopted accompanied us just for the analysis and design phase of creating the monitoring and adaptation layers for the distributed CCIM, and is limited to verifying the correct behaviour of the monitors and adaptors given the expected requirements. 

\begin{lstlisting}[frame=single, language=AsmetaL, basicstyle=\scriptsize, breaklines=true, caption={Verification of AsmetaSMV properties}, label={lst:prop_ver}]
> NuSMV -dynamic -coi -quiet C:\Work\Specs\ASMeta_Specs\code\Verification\SingleModelVerification.smv
-- specification ((AG ((!heartbeat_timeout(monitor_2) & (monitor_state(monitor_2) = 
WAIT_FOR_RESPONSE & heartbeat_response_arrived(monitor_2))) -> 
AX monitor_state(monitor_2) = COLLECT_DATA) & AG ((!heartbeat_timeout(monitor_1) &
(heartbeat_response_arrived(monitor_1) & monitor_state(monitor_1) = WAIT_FOR_RESPONSE)) 
-> AX monitor_state(monitor_1) = COLLECT_DATA)) & AG ((!heartbeat_timeout(monitor_3) 
& (monitor_state(monitor_3) = WAIT_FOR_RESPONSE & heartbeat_response_arrived(monitor_3
))) -> AX monitor_state(monitor_3) = COLLECT_DATA))  is true
-- specification ((AG (trigger_gossip(monitor_2) -> EF leader_state(leader_1) = EVALUATE) 
& AG (trigger_gossip(monitor_3) -> EF leader_state(leader_1) = EVALUATE)) 
& AG (trigger_gossip(monitor_1) -> EF leader_state(leader_1) = EVALUATE))  is true
-- specification ((AG (is_problem_discovered(monitor_1) -> AX monitor_state(monitor_1) = 
REPORT_PROBLEM) & AG (is_problem_discovered(monitor_2) -> AX 
monitor_state(monitor_2) = REPORT_PROBLEM)) & AG (is_problem_discovered(monitor_3) 
-> AX monitor_state(monitor_3) = REPORT_PROBLEM))  is true
-- specification ((AG ((monitor_state(monitor_1) = ASSIGN_DIAGNOSIS 
& !is_problem_discovered(monitor_1)) -> EX monitor_state(monitor_1) = LOG_DATA) 
& AG ((monitor_state(monitor_3) = ASSIGN_DIAGNOSIS & !is_problem_discovered(monitor_3)) 
-> EX monitor_state(monitor_3) = LOG_DATA)) & AG ((monitor_state(monitor_2) = 
ASSIGN_DIAGNOSIS & !is_problem_discovered(monitor_2)) -> EX monitor_state(monitor_2) 
= LOG_DATA))  is true
-- specification AG (leader_state(leader_1) = IDLE_LEADER ->
AX ((critical_diagnoses(leader_1) = 0 & failed_diagnoses(leader_1) = 0) 
& normal_diagnoses(leader_1) = 0))  is true
-- specification AG (leader_state(leader_1) = ASSESS -> AX (failed_diagnoses(leader_1) 
+ critical_diagnoses(leader_1)) + normal_diagnoses(leader_1) = 3)  is true
\end{lstlisting}

Modelling cloud systems has been proposed by the MODACloud project in order to obtain self-adaptive multi-cloud applications~\cite{modaclouds2016}. The project relies on CloudML\footnote{http://cloudml.org/} language for modelling the runtime processes and specifying the data, QoS models and monitoring operation rules~\cite{Bergmayr2015}. Resulting specifications are called models@runtime and allow bidirectional adaptation of models and execution based on updates performed at any of the both sides.

mOSAIC uses a Service Oriented Architecture (SOA) approach for cloud-based applications described by~\cite{Petcu2010}. It contains a brokering agent at the middleware level, dealing with maintaining promised SLAs and Key Performance Indicators (KPIs) at the infrastructure and application level. Reservoir cloud federation relies on Lattice framework for monitoring services~\cite{Rochwerger2009}, which follows a publish-subscribe model.

The ASM formal method has been previously used for specifying the behaviour of adaptive systems. For instance, Arcaini et al. propose in~\cite{Arcaini2017}, an ASM model for analyzing MAPE-K loops of self-adapting systems that follow a decentralized architecture. Flexibility and robustness to silent node failures of the specification is validated and verified with the aid of the ASMETA toolset. MAPE-K loops are important also for understanding the monitoring processes and their role for ensuring self-adapting systems. Ma et al. introduced the notion of Abstract State Services based on ASMs and described it for a flight booking over a cloud service case study~\cite{Ma2009}. 

Formal modelling was also used for specifying grids. ASM contributed to the description of job management and service execution in~\cite{Bianchi2011} and, as a further extension, in~\cite{Bianchi2013}. Specification of grids in terms of ASMs has been proposed also by~\cite{Nemeth2003}, with a focus in underlining differences between grids and normal distributed systems.

\section{Conclusion}
\label{sec:conclusion}
The advances of distributed systems aim to respond to ever growing requirements of clients. However, existing cloud and multi-cloud solutions propose services focused on the providers, rather than on the clients. For this reason, adoption of cloud services has been hindered. Shifting the focus towards a client-oriented platform for multi-clouds is, therefore, the goal of our research work. The premise was a middleware enhanced with security, privacy, adaptivity, and QoS measurements. 

In this chapter, we described a distributed version of the middleware and detailed the monitoring and adaptation layers, which complement and ensure a reliable execution. The highlighted workflow of the processes was consolidated with ASM specification models which capture the intended requirements of the system. Through thorough analysis of the model, we can identify design flaws, that otherwise, would propagate further to the implementation phase of the software development. Thus, the models were subject to validation by different scenarios and verification of specific meta-properties that reflect their quality. Properties of the solution are expressed in terms of CTL formulas and verified on the NuSMV version of the models.

In the next steps of our work, we aim to enrich the model with the communication between the monitors and the adaptation component. The monitors should be able to submit data related to an issue to the adapters, which in return should request an evaluation of the system after the enactment of an adaptation plan. We also propose building a weighted diagnosis, using the confidence degree of a monitor as its weight. In this way, monitors with a lower confidence degree will have a smaller contribution to the final evaluation. The extensions will be added to the current versions of the ASM models.


\end{document}